\documentclass[aps,pra,reprint,superscriptaddress,floatfix]{revtex4-1}
\usepackage{braket}
\usepackage{bm}
\usepackage{amsmath}
\usepackage{amsfonts}
\usepackage{graphicx}
\usepackage{placeins}
\usepackage{xcolor}
\graphicspath{{./figures/}}
\usepackage[]{hyperref}
\hypersetup{
    pdftitle={Optimal and Secure  Measurement Protocols for Quantum Sensor Networks},
    pdfauthor={Zachary Eldredge},
    bookmarksnumbered=true,     
    bookmarksopen=true,         
    bookmarksopenlevel=1,      
    colorlinks=true,            
    pdfstartview=Fit,           
    pdfpagemode=UseOutlines,    
    pdfpagelayout=TwoPageRight
}

\newcommand{\dd}[1]{\mathrm{d} #1}
\newcommand{\abs}[1]{\ensuremath{\left| #1 \right|}}
\newcommand{\half}{\frac{1}{2}}
\newcommand{\pdpd}[2]{\frac{\partial #1}{\partial #2}}
\newcommand{\expect}[1]{\left\langle #1 \right\rangle}

\newcommand{\sgn}[1]{\mathrm{sgn} \left( #1 \right)}

\newcommand{\adotq}{\bm{\alpha} \cdot \bm{\theta}}

\DeclareMathOperator{\Var}{Var}

\begin{document}
\title{Optimal and Secure  Measurement Protocols for Quantum Sensor Networks}
\date{\today}
\author{Zachary Eldredge}
\affiliation{Joint Quantum Institute, NIST/University of Maryland, College Park, MD 20742, USA}
\affiliation{Joint Center for Quantum Information and Computer Science, NIST/University of Maryland, College Park, MD 20742, USA}
\author{Michael Foss-Feig}
\affiliation{Joint Quantum Institute, NIST/University of Maryland, College Park, MD 20742, USA}
\affiliation{Joint Center for Quantum Information and Computer Science, NIST/University of Maryland, College Park, MD 20742, USA}
\affiliation{United States Army Research Laboratory, Adelphi, MD 20783, USA}
\author{Jonathan A. Gross}
\affiliation{Center for Quantum Information and Control, University of New Mexico, Albuquerque, NM 87131, USA}
\author{S. L. Rolston}
\affiliation{Joint Quantum Institute, NIST/University of Maryland, College Park, MD 20742, USA}
\author{Alexey V. Gorshkov}
\affiliation{Joint Quantum Institute, NIST/University of Maryland, College Park, MD 20742, USA}
\affiliation{Joint Center for Quantum Information and Computer Science, NIST/University of Maryland, College Park, MD 20742, USA}

\begin{abstract}
	Studies of quantum metrology have shown that the use of many-body entangled states can lead to an enhancement in sensitivity when compared to unentangled states. In this paper, we quantify the metrological advantage of entanglement in a setting where the measured quantity is a linear function of parameters individually coupled to each qubit. We first generalize the Heisenberg limit to the measurement of non-local observables in a quantum network, deriving a bound based on the multi-parameter quantum Fisher information.  We then propose measurement protocols that can make use of Greenberger-Horne-Zeilinger (GHZ) states or spin-squeezed states and show that in the case of GHZ states the protocol is optimal, i.e., it saturates our bound. We also identify nanoscale magnetic resonance imaging as a promising setting for this technology.
\end{abstract}

\maketitle

\section{Introduction}

Entanglement is a valuable resource in precision measurement, as measurements using entangled probe systems have fundamentally higher optimal sensitivity than those using unentangled states \cite{Bollinger1996}. A generic measurement using $N$ unentangled probes will have a standard deviation from the true value asymptotically proportional to $1/ \sqrt{N}$. By using $N$  maximally entangled probes, a single parameter coupled independently to each probe system can be measured with an uncertainty proportional to $1/N$. This is the best possible scaling consistent with the Heisenberg uncertainty principle and is known as the Heisenberg limit \cite{Wineland1992,Bollinger1996}.
The procedure can also be reversed--enhanced sensitivity to disturbances can provide experimental evidence of entanglement \cite{Pezze2009,Strobel2014,Toth2012}. 

\begin{figure}[tb]
	\includegraphics[width=8.6cm]{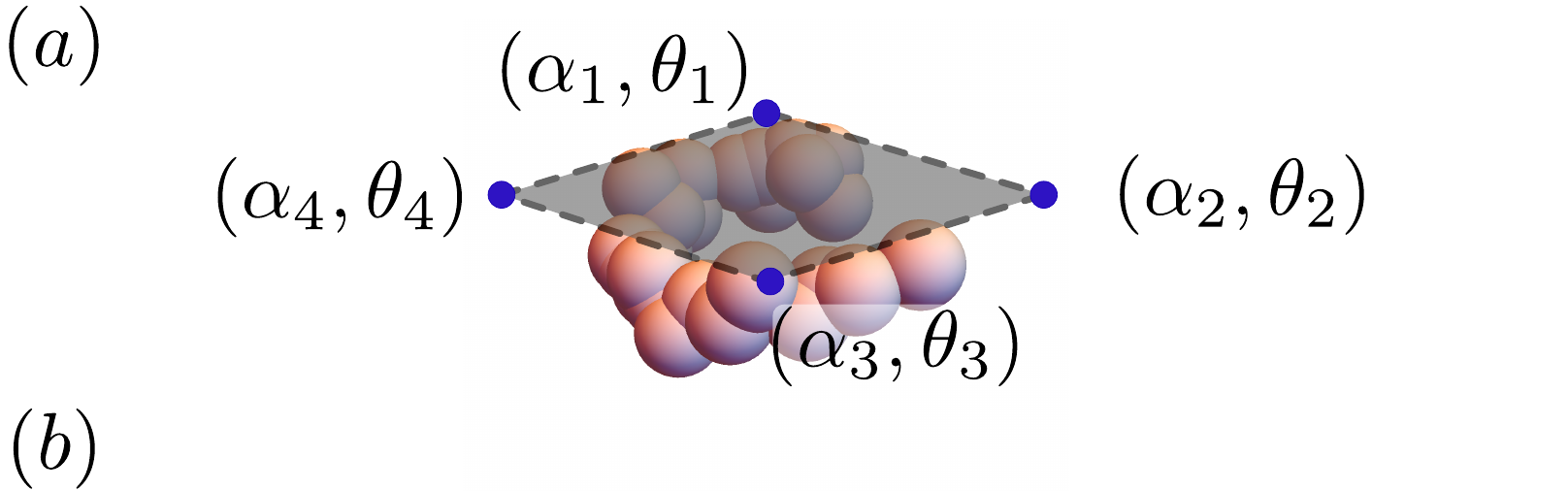}
	\includegraphics[width=8.6cm]{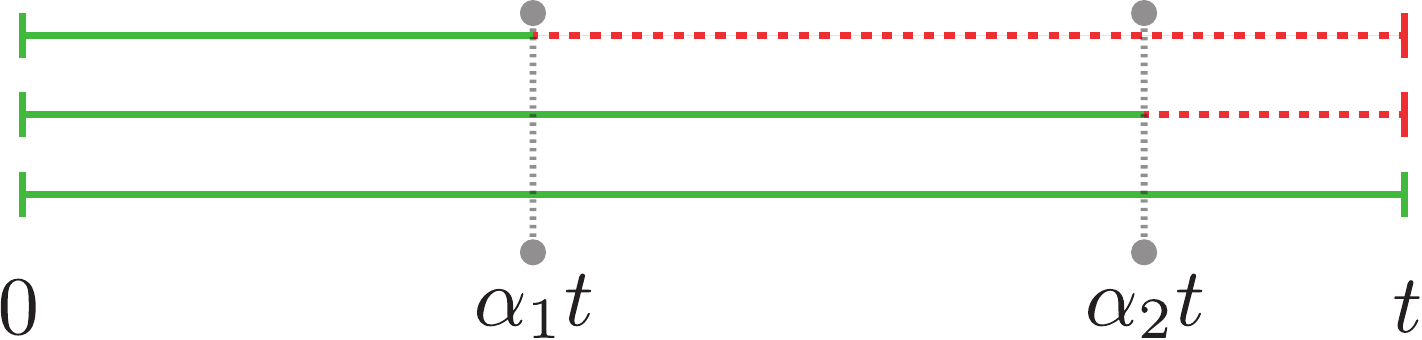}
	\caption{(a) An illustration of the network setup in a nanoscale NMR  setting. Nodes, located at different points relative to a large molecule, share an entangled state; at each node there is both an unknown parameter $\theta_i$ and a known relative weight $\alpha_i$. We are concerned with estimating $\adotq$. (b) Illustration of the partial time evolution protocol for three qubits. Solid green segments of the timeline represent periods when a qubit is evolving due to coupling to the local parameter $\theta_i$, while dashed red segments represent periods after the qubit stops evolving. The switches occur at times corresponding to the qubits' weights in the final linear combination. The weight of the last qubit is $\alpha_3 = 1$.}
	\label{fig:setup}
\end{figure}

Measurements making use of entanglement usually couple one parameter to $N$ different systems \cite{Bollinger1996,Dinani2016,Kessler2014}. However, the emerging potential of long-range quantum information opens new avenues for metrology \cite{Komar2014,Komar2016} and entanglement distribution \cite{Valivarthi2016}. The ability to distribute entanglement across spatially separated regions has already been used for recent loophole-free tests of Bell's inequality \cite{Giustina2015b,Shalm2015,Hensen2015}.
In this work, we are interested in coupling $N$ parameters to $N$ different systems, which may be spatially separated, and measuring a linear function of all of them (see Fig.~\ref{fig:setup}a) such as a single mode of a spatially varying field. Such measurements may be of interest in geodesy, geophysics, or medical imaging \cite{Nabighian2005,Ding2013,Wright2004,Stacey1964,Hamalainen1993}, but in this paper we focus on potential application to nanoscale nuclear magnetic resonance (NMR) imaging. Later in this paper we will discuss precisely how our method might apply in this setting.

The function $q$ we wish to measure is a weighted sum of the deterministic individual parameters $\theta_i$, where $i$ indexes the individual systems and each weight is denoted by a known real number $\alpha_i$, 
\begin{equation}
	q = \sum_{i=1}^N \alpha_i \theta_i = \adotq.
	\label{eqn:qdef}
\end{equation}
In this paper, we characterize the advantage entanglement provides in this setting and construct an optimal strategy equivalent to turning some qubits' evolution ``on'' and ``off'' for time proportional to the weight with which their parameter contributes to the function $q$ (see Fig.~\ref{fig:setup}b). With this scheme of ``partial time evolution,'' we can measure a linear function with the minimum variance permitted by quantum mechanics, which can be viewed as an extension of the Heisenberg limit to linear combinations. We will also show that our method can protect the secrecy of the result, allowing the network as a whole to perform a measurement without eavesdroppers learning any details of $\adotq$.

\section{Setup}
We consider a system in which there are $N$ sensor nodes. Each sensor node $i$ possesses a single qubit coupled to an unknown parameter $\theta_i$ unique to each node. We suppose that the state evolves unitarily under the Hamiltonian
\begin{equation}
	\label{eqn:hamiltonian}
	\hat{H} = \hat{H}_c (t) + \sum_{i=1}^N \half \theta_i \hat{\sigma}^z_i.
\end{equation}
Here, $\hat{H}_c (t)$ is a time-dependent control Hamiltonian chosen by us, which may include coupling to additional ancilla qubits and $\hat{\sigma}^{x,y,z}_i$ are the Pauli operators acting on qubit $i$. 
We wish to measure the quantity $q$ defined in Eq.~\eqref{eqn:qdef}.
We assume that $\forall i : \abs{\alpha_i} \leq 1$ and additionally that there is at least one $\alpha_i$ such that $\alpha_i = 1$. These conditions simply set a scale for the function, and for an arbitrary $\bm{\alpha}$ all that is needed is division by the largest $\alpha_i$ to meet this requirement. As an example, a network with two nodes interested in measuring the contrast between those nodes would set $\bm{\alpha} = \left( 1, -1\right)$ to measure $\theta_1 - \theta_2$. We would like to establish how well an arbitrary measurement of $\adotq$ can be made and what the best measurement protocol is for doing so. By ``protocol'' we mean three different choices: (1) which input state we begin with, (2) what auxiliary control Hamiltonian $\hat{H}_c(t)$ we implement, and (3) how the final measurement is made. 

We define the quality of measurement in terms of an estimator, $Q$, constructed from experimental data. (Throughout this paper, we denote operators with hats, vectors by boldface, quantities to be estimated by lowercase, and corresponding estimators by uppercase.) We assume that the estimator is unbiased, so that its expectation value is the true value $\mathbf{E} \left[Q \right] = q$. Then our metric for the quality of the measurement is the average squared error, or variance, of the estimator,
\begin{equation}
	\Var Q = \mathbf{E} \left[ \left( Q - q \right)^2  \right].
\end{equation}
If measurements of $\theta_i$ can be made locally with accuracy $\Var \Theta_i$ for an estimator $\Theta_i$, then we could compute the linear combination by local measurements and classical computation.  In this case, the variance is given by classical statistical theory as  $\Var Q = \| \bm{\alpha} \|^2  \Var \Theta_0$ assuming that $\Var \Theta_i$ is identical at each site and equal to $\Var \Theta_0$. A measurement of an individual $\theta_i$ in Eq.~\eqref{eqn:hamiltonian} can be made in time $t$ with a variance of $1/t^2$ \cite{Wineland1992}. Therefore, our entanglement-free figure of merit is
\begin{equation}
	\Var Q \geq \frac{\| \bm{\alpha} \|^2}{t^2}.
	\label{eqn:sql}
\end{equation}
We consider this the standard quantum limit for networks. To compare to the typical case where $N$ independent qubits measure a single parameter, consider the average $\bar{\theta}$, which is equivalent to setting all $\alpha_i = 1$ and then using $\bar{\Theta} = Q/N$ to obtain $\Var \bar{\Theta} = 1/ N t^2$.  It is our goal in this paper to present a means to improve on the limit in Eq.~\eqref{eqn:sql}.

\section{Heisenberg Limit for Sensor Networks }
\subsection{Using Fisher Information Matrix}
Our task is to perform parameter estimation on a quantum system evolving under some set of parameters $\{\theta_i\}$ linearly coupled to sensor qubits as in Eq.~\eqref{eqn:hamiltonian} \cite{Braunstein1996,Paris2008,Boixo2007,Pang2014}. Although we are only interested in measuring a single number, we still need to treat a system that has many parameters in the evolution, necessitating the use of a multi-parameter theory as in Refs.~\cite{Baumgratz2015,Crowley2014,Gao2014,Genoni2013,Humphreys2013,Kok2017,Yue2014,Zhang2014, Proctor2017}. It is known from classical estimation theory that, given a probability distribution $p(z)$ over a set of outcomes $z$ that depends on a number of parameters, all estimators of the parameters obey the Cram\'{e}r-Rao inequality \cite{Helstrom1968,Helstrom1969},
\begin{equation}
	\Sigma \geq \frac{F^{-1}}{M}.
	\label{eqn:crb}
\end{equation}
Here, $M$ is the number of experiments performed, $F$ is the Fisher information matrix (see below), and $\Sigma$ is the covariance matrix, where $\Sigma_{ij} = \mathbf{E} \left[ \left( \Theta_i  - \theta_i \right) \left( \Theta_j  - \theta_j \right) \right]$. The inequality is a matrix inequality, meaning that $M \Sigma - F^{-1}$ is positive semidefinite. We will concern ourselves with the single-shot Fisher information, and set $M = 1$ from now on. The Fisher information matrix captures how each parameter changes the probability distribution of outcomes,
\begin{equation}
	F_{ij} = \int p(z) \left( \pdpd{ \ln p(z)}{\theta_i} \right)  \left( \pdpd{ \ln p(z)}{\theta_j} \right) \dd{z}.
	\label{eqn:fishermat}
\end{equation}
This bound is a purely classical statement about probability distributions, and is saturated asymptotically using a maximum-likelihood estimator \cite{Braunstein1994}. Note that although we have presented the formulas for the Fisher information matrix, in the case of a single parameter the Fisher information will be a scalar which can be obtained by setting $i = j$ in Eq.~\eqref{eqn:fishermat}. 

Quantum theory bounds the probability distributions that can result from a state evolved under a parameter-dependent unitary operation \cite{Braunstein1996}. We thus define the quantum Fisher information $F_Q$ for a process with a given initial state as the maximization of the Fisher information over all possible measurement schemes. This gives rise to the quantum Cram\'{e}r-Rao bound (QCRB), which simply replaces $F$ with $F_Q$ in Eq.~\eqref{eqn:crb}. A matrix element of $F_Q$ for a pure state evolving under a Hamiltonian $\hat{H}$ is given by
\begin{equation}
	\left( F_{Q} \right)_{ij} = 4 t^2 \left[ \expect{\hat{g}_i \hat{g}_j} - \expect{\hat{g}_i} \expect{\hat{g}_j} \right],
	\label{eqn:qfishmat}
\end{equation}
where $\hat{g}_i =  \left( \partial \hat{H} / \partial \theta_i \right)$ is the generator corresponding to parameter $i$. For instance, in Eq.~\eqref{eqn:hamiltonian} the generator $\hat{g}_i$ is the operator $\half \hat{\sigma}^z_i$. Unlike the Cram\'{e}r-Rao bound, the QCRB cannot always be satisfied, even asymptotically. However, in the setting of this paper, where all generators commute, it can be \cite{Baumgratz2015}. Equation~\eqref{eqn:crb} then takes the form:
\begin{equation}
	\Sigma \geq F^{-1} \geq F_Q^{-1}.
	\label{eqn:crbchain}
\end{equation}

To formulate the appropriate Cram\'{e}r-Rao bound in the case where the quantity we wish to estimate is a linear combination of the $\theta_i$, we simply use the fact that the variance of a linear combination $\adotq$ can be written as $\bm{\alpha}^T \Sigma \bm{\alpha}$. It follows immediately from Eq.~\eqref{eqn:crbchain} that
\begin{equation}
	\Var Q \geq \bm{\alpha}^T F_Q^{-1} \bm{\alpha}.
	\label{eqn:lincombcrb}
\end{equation}
Note that although we began by considering the full covariance matrix, we now focus on just a \emph{single} scalar $\bm{\alpha}^T F_Q^{-1} \bm{\alpha}$ because our quantity of interest is a \emph{single} linear transformation of the original parameters.

In order to properly define the Cram\'{e}r-Rao bound, it is necessary to consider the fact that $F$ and $F_Q$ are only positive semi-definite and not necessarily invertible. For instance, if a parameter has no effect on probabilities at all, then it cannot be estimated from experimental results and the bound on the variance of its estimator is undefined. To sidestep this issue, we can instead look at $\tilde{F}_Q$, the quantum Fisher information projected onto its own image \cite{Proctor2017}, assuming that $\bm{\alpha}$ has no overlap with the kernel of $F_Q$.  This matrix (and its inverse) are now both positive definite, meaning they can always be inverted. Equation~\eqref{eqn:lincombcrb} is therefore always well-defined if $\tilde{F}_Q$ is used.

Since $\tilde{F}_Q$ is Hermitian and positive definite, $\sqrt{\tilde{F}_Q}$ is Hermitian. We can then write the following for an arbitrary real $\bm{b}$ by invoking the Cauchy-Schwarz inequality:
\begin{align}
	\bm{\alpha}^T \tilde{F}_Q^{-1} \bm{\alpha}   &= \frac{\| \sqrt{\tilde{F}_Q^{-1}} \bm{\alpha} \|^2 \| \sqrt{\tilde{F}_Q} \bm{b} \|^2}{ \bm{b}^T \tilde{F}_Q \bm{b} } \\ &\geq \frac{\| \bm{\alpha}^T \sqrt{\tilde{F}_Q^{-1}} \sqrt{\tilde{F}_Q} \bm{b} \|^2}{ \bm{b}^T \tilde{F}_Q \bm{b} } \\
	& \geq \frac{\| \bm{\alpha}^T \bm{b} \|^2}{ \bm{b}^T \tilde{F}_Q \bm{b} }.
\end{align}
Taking $\bm{b}$ to be the $b$th element of the standard basis gives
\begin{equation}
	\Var{Q} \geq \bm{\alpha}^T \tilde{F}_Q^{-1} \bm{\alpha} \geq \frac{ \alpha_b^2 }{ \left(\tilde{F}_Q \right)_{bb}}.
	\label{eqn:multiparamqcrb}
\end{equation}
Here, $\left( \tilde{F}_Q \right)_{bb}$ is the quantum Fisher information for a single parameter, as defined by Eq.~\eqref{eqn:qfishmat}. In Ref.~\cite{Boixo2007}, it was shown that for any time-dependent control Hamiltonian $\hat{H}_c(t)$, including those with ancilla qubits, 
\begin{equation}
	\left(\tilde{F}_Q \right)_{bb} \leq t^2 \| \hat{g}_b \|_s^2. 
\end{equation}
Here $\| \hat{g}_b \|_s$ is the operator seminorm (difference between the largest and smallest eigenvalues) of the generator corresponding to parameter $\theta_b$. Our final bound comes from applying this condition and recognizing that the formula must hold for all $b$:
\begin{equation}
	\Var Q \geq  \max_b \frac{ \alpha_b^2}{t^2 \| \hat{g}_b \|_s^2}.
	\label{eqn:generalbound}
\end{equation}
We emphasize that Eq.~\eqref{eqn:generalbound} remains true no matter what time-dependent control $\hat{H}_c(t)$ is applied. 

In Eq.~\eqref{eqn:hamiltonian}, all $\hat{g}_b = \half \hat{\sigma}^z_b$, $\| \hat{g}_b \|_s = 1$, and we find a bound, 
\begin{equation}
	\Var Q \geq  \max_i \frac{\alpha_i^2}{t^2} = \frac{1}{t^2}.
	\label{eqn:ptebound}
\end{equation}
Here we have used the fact that the largest $\alpha_i = 1$. If we want to estimate the average of the $\theta_i$, then all qubits are equally weighted and the desired quantity is $\bar{\theta} = q/N$, so $\Var{\bar{\Theta}} \geq 1 / N^2 t^2$ and we reproduce the desired Heisenberg scaling which is more precise than the $1/N$ in Eq.~\eqref{eqn:sql}. However, note that if we wanted to estimate only a single $\theta_i$, then we would not benefit from the entanglement. In general, we can, for some situations, greatly improve the precision of parameter esitmation with nonlocal techniques if the parameter itself is also non-local. Our bound allows us to explore the full range of possible $\bm{\alpha}$ between these two extremes. Compared to the bound on unentangled states [Eq.~\eqref{eqn:sql}], Eq.~\eqref{eqn:generalbound} simply picks out the largest contribution due to uncertainty from a single site. Equation~\eqref{eqn:generalbound} can be viewed as an extension of the usual Heisenberg bound to linear combinations of parameters.

\begin{figure}[tb]
	\includegraphics[width=8.6cm]{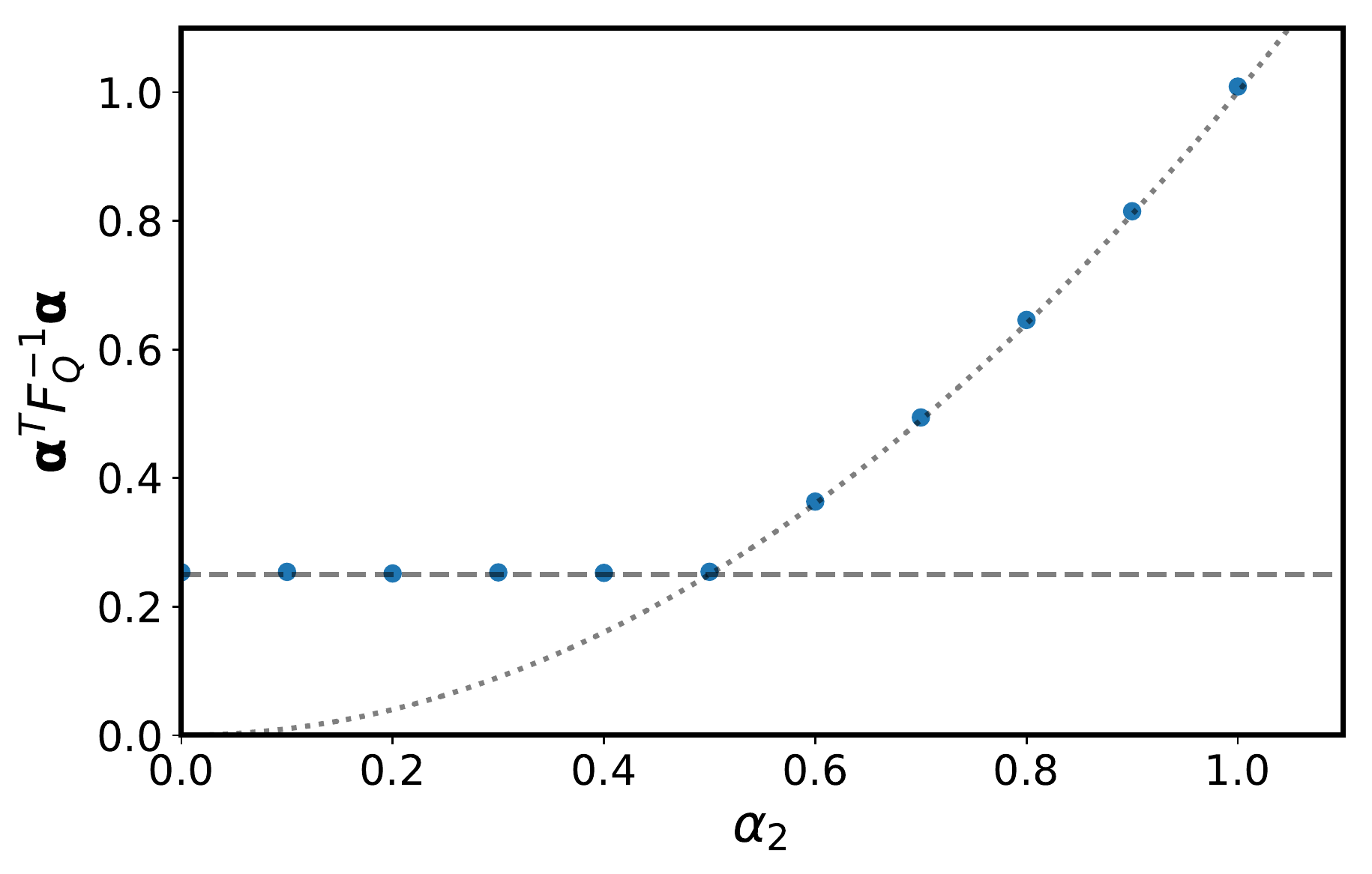}
	\caption{Numerical optimization of $\bm{\alpha}^T F^{-1}_Q \bm{\alpha}$ for two qubits with $\alpha_1 = 1$ compared to the bound predicted by our analytic result. Each point is generated by running a gradient descent algorithm until convergence; the control parameters begin at small random values. The dashed (dotted) line is the analytic bound derived from the first (second) qubit. As $\alpha_2$ increases, the second qubit becomes the source of the relevant bound.}
	\label{fig:graddescent}
\end{figure}

We can illustrate the above argument by optimizing over the space of all control Hamiltonians $\hat{H}_c(t)$. As this is computationally expensive, we limit ourselves to a two-qubit sensor network with no ancillas. The Hamiltonians we optimize over include enough operators to provide universal control on two qubits, meaning we can effectively modify the input state as well as the final measurement basis in order to optimize the Fisher information. In order to test the form of our bound, Eq.~\eqref{eqn:generalbound}, which depends both on relative weights of each parameter and the underlying generator, we couple $\theta_1$ to a generator $\hat{\sigma}^z_1$ which has $\| \hat{\sigma}^z_1 \|_s = 2$. We leave the second qubit coupled to a generator $\half \hat{\sigma}^z_2$ as in Eq.~\eqref{eqn:hamiltonian}. The bound corresponding to the first qubit from Eq.~\eqref{eqn:generalbound} is $\alpha_1^2/4 t^2$ and that of the second qubit is $\alpha_2^2 / t^2$. In our numerics, we set $\alpha_1 = t = 1$, meaning the two bounds are $1/4$ and $\alpha_2^2$. Our analytic result leads us to believe therefore that if $\alpha_2^2 > 1/4$, the minimum possible variance should be $\alpha_2^2$. However, if $\alpha_2^2 < 1/4$, then the lower bound should be $1/4$. That behavior is precisely what we find through the numerical optimization shown in Fig.~\ref{fig:graddescent}, confirming Eq.~\eqref{eqn:generalbound}.

\subsection{Using Single-Parameter Bounds}
\label{sec:spbound}
It is tempting to dismiss the above argument as unnecessarily complicated, as the ultimate quantity of interest is only a single parameter. Why not simply apply the Cram\'{e}r-Rao bound directly to $\adotq$ instead of using the matrix approach? We will now show that the single-parameter bound that arises from naive application of the Cram\'{e}r-Rao bound is looser than Eq.~\eqref{eqn:generalbound}. This gap occurs because the single-parameter bound can only be applied if there is only one unknown parameter controlling the evolution of the input state, which implicitly places a constraint on the other components of the field. Later, we will discuss how the single-parameter approach can be amended to take this into account and agree with Eq.~\eqref{eqn:generalbound}.

To apply the single-parameter Cram\'{e}r-Rao bound to our evolution Hamiltonian Eq.~\eqref{eqn:hamiltonian}, we consider the Hamiltonian as $\half\bm{\theta} \cdot \bm{\hat{\sigma}}$ where $\bm{\hat{\sigma}}$ is simply a vector of operators whose $i$th element is $\hat{\sigma}^z_i$. We then rewrite $\bm{\theta}$ in a new basis,
\begin{equation}
	\bm{\theta} = \sum_{i=0}^{N - 1} \left(\bm{\alpha}_i \cdot \bm{\theta} \right) \bm{\beta}_i.
	\label{eqn:rebasis}
\end{equation}
We assume that $\bm{\alpha}_0 = \bm{\alpha}$ and that the other $\bm{\alpha}_{i>0}$ make up a basis. The set of vectors $\bm{\beta}_i$ is then a dual basis such that $\bm{\alpha}_i \cdot \bm{\beta}_j = \delta_{ij}$. (For this basis as well, we will drop the subscript 0 to indicate that this particular vector corresponds to the parameter of interest.) The advantage of rewriting $\bm{\theta}$ this way is that we can now identify the term in the Hamiltonian which is proportional only to $\bm{\alpha} \cdot \bm{\theta}$. The generator corresponding to the quantity $\adotq$ is:
\begin{equation}
	\hat{g} = \pdpd{\hat{H}}{\left( \adotq \right)} = \frac{\bm{\beta} \cdot \bm{\hat{\sigma}}}{ 2}.
\end{equation}

To obtain the quantum Fisher information corresponding to this generator, we consider the variance of the operator $\hat{g}$. The maximum variance of this generator is given by the operator seminorm \cite{Boixo2007}. Using this fact, we can write:
\begin{equation}
	F_Q \leq t^2  \| \hat{g} \|_s^2  =   t^2 \left( \sum_i \abs{\beta_i} \right)^2.
	\label{eqn:badbound}
\end{equation}

In general the bound on $\Var Q$ derived from Eq.~\eqref{eqn:badbound} is a looser lower bound than Eq.~\eqref{eqn:ptebound}. For example, with $\bm{\alpha} = (1, \half)$ and $\bm{\alpha}_1 = (\half, -1)$, this implies that $\bm{\beta} = (\frac{4}{5}, \frac{2}{5})$. Equation~\eqref{eqn:badbound} would suggest that
\begin{equation}
	\Var Q \geq \frac{25}{36 t^2},
\end{equation}
which is looser than the $1/t^2$ given by Eq.~\eqref{eqn:generalbound}. This discrepancy can be addressed by thinking more closely about the process of choosing a new basis. We will use the seminorm condition again to bound the maximum possible Fisher information. To start calculating the seminorm, we express it in terms of the elements of $\bm{\beta}$:
\begin{equation}
	\| \hat{g} \|_s = \| \sum_{j} \beta_{j} \half \hat{\sigma}^z_j \|_s = \sum_j \abs{\beta_{j}}.
	\label{eqn:genbound}
\end{equation}
We will now show that it is possible to choose a basis such that the seminorm in Eq.~\eqref{eqn:genbound} goes to infinity. This shows that the approach which led us to Eq.~\eqref{eqn:badbound} should not be applied blindly, and we will then discuss how to control for this issue. First, an illustration of the bound diverging. Suppose that in a two-parameter problem, the basis vectors we choose are $\bm{\alpha}$ and $\bm{\alpha}'$.  It can then be shown by direct computation of the matrix inverse that yields the dual basis that the implied maximum Fisher information from Eq.~\eqref{eqn:genbound} is:
\begin{equation}
	F \leq \| \hat{g} \|_s = \frac{\alpha'_2 + \alpha'_1}{\abs{ \alpha_1 \alpha'_2 - \alpha'_1 \alpha_2}}.
\end{equation}
If we then choose $\bm{\alpha}' = \left(\alpha_1/\alpha_2 + \varepsilon, 1 \right)$, it follows that:
\begin{equation}
	F \leq \frac{1 + \varepsilon + \frac{\alpha_1}{\alpha_2}}{\varepsilon \alpha_2}.
\end{equation}
As $\varepsilon \to 0$, this becomes arbitrarily large. From this we conclude that our previous approach was ill-advised as it can yield arbitrarily small lower bounds on the estimator variance -- using this basis, we would conclude that the right-hand side of Eq.~\eqref{eqn:badbound} could be $\infty$.

In order to produce a useful bound from Eq.~\eqref{eqn:genbound}, we recognize that any possible choice of basis must yield a valid bound. Therefore, rather than look at one particular basis (as we did in deriving Eq.~\eqref{eqn:badbound}) we instead need to optimize for the highest lower bound over all possible choices of basis. Finding the tightest bound on Fisher information will then produce the highest lower bound on parameter uncertainty. To do this, we first write the following chain of inequalities using the relationship of $\bm{\alpha}$ and $\bm{\beta}$:
\begin{align}
	1 &= \sum_j \alpha_{j} \beta_j \leq
	\sum_j \abs{ \alpha_j  \beta_j} \leq 
	\sum_j \abs{ \beta_j}, 
	\label{eqn:matrixbound}
\end{align}
where the last line follows due to the fact that $\abs{\alpha_j} \leq 1$. Note that we can achieve equality,
\begin{equation}
	\sum_j \abs{\beta_j} = 1,
\end{equation}
by taking the other $N-1$ basis vector $\bm{\alpha}_j$ to be unit vectors $\bm{e}_j$ in the standard basis, making sure that the $j$ that does not appear has $\alpha_j \neq 0$ to ensure the entire space is spanned. Now we look to the minimum possible value of $\|\hat{g}\|_s$. The minimum possible value is interesting to us because the minimum $\|\hat{g} \|_s$ will be the choice of basis for which the bound on Fisher information is tightest.  

It follows from Eq.~\eqref{eqn:genbound} and Eq.~\eqref{eqn:matrixbound} that the minimum seminorm $\| \hat{g}\|_s$ is equal to 1, implying that the maximum value for $\Var \hat{g}$ is $1/4$ \cite{Boixo2007}. Using this to optimize the bound in Eq.~\eqref{eqn:badbound} over all possible choices of re-parameterization implies that $\Var \adotq \geq 1/t^2$, just as we found in Eq.~\eqref{eqn:ptebound}. 

The single-parameter bound \textit{is} applicable in our situation, but it requires careful accounting of the influence of other parameters in the problem. The reason that our previous results such as Eq.~\eqref{eqn:generalbound} do not hold in this case is that Cram\'{e}r-Rao bound does not apply if we can take advantage of constraints on the signal field $\bm{\theta}$ to improve our estimation strategy.

These naive single-parameter bounds can be applied and saturated if the field structure is known before the measurement takes place. To demonstrate, suppose that for a set of fields $\bm{\theta}$ where we wish to learn $\adotq$, we know that the fields are proportional to $\alpha_i$. Then we can write the total field as:
\begin{equation}
	\bm{\theta} = q \frac{\bm{\alpha}}{\| \bm{\alpha} \|^2},
\end{equation}
and our goal is to estimate $q = \adotq$.  This is now a truly one-parameter problem, enabling a new strategy which saturates Eq.~\eqref{eqn:badbound}.
By defining $\bm{w}$ as a new vector such that $w_i = \sgn{\alpha_i}$, we can measure the quantity
\begin{equation}
	\bm{w} \cdot \bm{\theta} = q \frac{ \sum \abs{\alpha_i}}{\| \bm{\alpha} \|^2}.
\end{equation}
Since $\bm{w}$ is a linear combination which satisfies the condition $\abs{w_i} \leq 1$, we can estimate $q' = \bm{w} \cdot \bm{\theta}$ with accuracy bounded by $1/t^2$ as shown in Sec.~\ref{sec:protocols}. Then:
\begin{align}
	\Var \left( Q' \right) &= \Var \left( Q \frac{ \sum \abs{\alpha_i}}{\| \bm{\alpha} \|^2} \right) \geq \frac{1}{t^2}, \\
	\implies \Var Q  &\geq \frac{\| \bm{\alpha} \|^4}{ t^2 \left(\sum \abs{\alpha_i} \right)^2}.
\end{align}
This saturates the bound in Eq.~\eqref{eqn:badbound}.
The reason we are able to outperform Eq.~\eqref{eqn:generalbound} is that we have assumed something about the structure of the field which reduces it to a lower-dimensional problem. This is only possible by using knowledge about components of $\bm{\theta}$ not parallel to $\bm{\alpha}$. Otherwise, there is no guarantee that $\bm{\theta}$ will be proportional to $\bm{\alpha}$. In general cases $\bm{w} \cdot \bm{\theta}$ will contain noise from ``undesired'' components.

In many situations where the field structure is known, new strategies can be introduced which may outperform our previous results, even asymptotically. Consider as an example a case with a field:
\begin{equation}
	\bm{\theta} = \theta \frac{\bm{\alpha}}{\| \bm{\alpha} \|^2} + \theta_\gamma \bm{\gamma},
\end{equation}
where $\bm{\alpha} \cdot \bm{\gamma} = 0$ and $\theta_\gamma$ is a nuisance parameter describing the field magnitude orthogonal to $\bm{\alpha}$. Any field can be written in this way to separate out the $\bm{\alpha}$ component. Suppose we measure $\bm{w} \cdot \bm{\theta}$. We know that:
\begin{equation}
	\Var Q' \geq \frac{1}{t^2}
\end{equation}
is an achievable bound. By writing $\bm{w} = c_\alpha \bm{\alpha} + c_\gamma \bm{\gamma}$, decomposing $\bm{w}$ into its only relevant components, we can obtain the following bound on $\Var Q$:
\begin{equation}
	\Var Q \geq \frac{1}{t^2 c_\alpha} + c_\gamma^2 \| \bm{\gamma} \|^2.
\end{equation}
Therefore, the optimal strategy is to pick a $\bm{w}$ which maximizes $\bm{w} \cdot \bm{\alpha}$ while minimizing (preferably to zero) $\bm{w} \cdot \bm{\gamma}$. However, in general, learning the structure of the field perpendicular to $\bm{\alpha}$ is just as difficult as learning the component parallel to $\bm{\alpha}$, so beginning from a state of ignorance, it is still optimal to measure $\adotq$ rather than a different linear combination. 

The bound in Eq.~\eqref{eqn:badbound} can actually be found by other statistical methods which fully treat the initial multi-parameter structure, for instance, the constrained Cram\'{e}r-Rao bound of Ref.~\cite{Stoica1998}. It can also be derived from the Van Trees inequality \cite{VanTrees2013} by assuming that we have pre-existing knowledge that the components of $\bm{\theta}$ perpendicular to $\bm{\alpha}$ have a normal distribution of width $\varepsilon$ and then taking the limit $\varepsilon \to 0$.

If rather than a constraint we simply have some initial information in the form of a prior distribution, the Van Trees inequality (which takes into account that prior information) will reduce to the Cram\'{e}r-Rao bound in the limit of many measurements. This is because the information gained from measurements scales linearly with the number of measurements while the prior information is static.
\section{Protocols}
\label{sec:protocols}
We now present two protocols that saturate the bound of Eq.~\eqref{eqn:ptebound} and are therefore optimal. The first begins from the conceptually simple Greenberger-Horne-Zeilinger (GHZ) state or a spin-squeezed state and uses time-dependent control during phase accumulation to produce an output state sensitive to the desired $\adotq$, while the second method uses a more complicated initial state but requires no control during the phase accumulation.

\subsection{Protocols Involving Time-Dependent Control}
\label{sec:tdprotocols}
\subsubsection{Using GHZ Input State}
\label{sec:pte}
We start by considering an $N$-qubit GHZ state:
\begin{equation}
	\frac{1}{\sqrt{2}} \left( \ket{0}^{\otimes N} + \ket{1}^{\otimes N} \right).
\end{equation}
Under $\hat{\sigma}^z$ evolution, each $\ket{1}$ accumulates a phase relative to $\ket{0}$. By allowing qubits to accumulate phase proportional to the desired weight $\alpha_i$, we obtain a final state in which $\ket{1}^{\otimes N}$ has accumulated a total phase of $\adotq t$ relative to $\ket{0}^{\otimes N}$. 
We refer to our protocol as ``partial time evolution'' because it relies on a qubit undergoing evolution for a fraction of the total measurement time (see Fig.~\ref{fig:setup}). We can realize this by applying $\hat{\sigma}^x_i$ to a qubit at time $t_i = t \left(1 + \alpha_i \right)/2$ so that the qubit evolution will be identical to evolving it for a time $\alpha_i t$. Note that if there is a fixed experimental time $t$, this scheme can realize values of $\alpha_i \in \left[ -1, 1 \right]$, which motivates our restrictions on the values of individual $\alpha_i$. Specifying this sequence of gates identifies the $\hat{H}_c(t)$ which defines the protocol.
The result of this protocol is an effective evolution according to the unitary operator
\begin{equation}
	\hat{U}(t) = e^{ - i \frac{t}{2} \sum_{i = 1}^N \alpha_i \theta_i \hat{\sigma}^z_i}. 
	\label{eqn:pteunitary}
\end{equation}
Under this evolution, the final state is:
\begin{align}
	\frac{1}{\sqrt{2}} \left( e^{-i \frac{t}{2} q} \ket{0}^{\otimes N} + e^{i \frac{t}{2} q }  \ket{1}^{\otimes N} \right).
	\label{eqn:postptestate}
\end{align}
Now we make a measurement of the overall parity of the state, $ \hat{P} = \bigotimes_{i=1}^N \hat{\sigma}^x_i$. The details of this measurement and calculation of $\langle \hat{P} \rangle$ are given in Ref.~\cite{Bollinger1996}; notably, the measurement can be performed locally at each site.
Measurement of the time-dependent expectation value $\langle \hat{P} \rangle (t)$ allows for the estimation of $Q$ with accuracy \cite{Wineland1994}
\begin{equation}
	\Var Q = \frac{\Var \hat{P}(t)}{\left(\partial{\expect{\hat{P}}} / \partial{q} \right)^2} = \frac{ \sin^2 q t }{t^2 \sin^2 q t } = \frac{1}{t^2},
	\label{eqn:ptesensitivity}
\end{equation}
saturating the bound in Eq.~\eqref{eqn:ptebound} and Fig.~\ref{fig:graddescent}.

We can also directly evaluate $F_Q$ and $F$ for this protocol. $F_Q$ can be found by noting that this protocol is identical to evolution under the Hamiltonian $\half \sum \alpha_i \theta_i \hat{\sigma}_i^z$. Therefore the quantum Fisher information matrix $F_Q$ is simply
\begin{equation}
	\left( F_Q \right)_{ij} = t^2 \left[ \expect{ \alpha_i \hat{\sigma}_i^z \alpha_j \hat{\sigma}_j^z } - \expect{ \alpha_i \hat{\sigma}_i^z } \expect{ \alpha_j \hat{\sigma}_j^z } \right] = \alpha_i \alpha_j t^2.
\end{equation}
Furthermore, we can show that this $F_Q$ satisfies the second inequality in Eq.~\eqref{eqn:multiparamqcrb}. The inverse of $\tilde{F}_Q$ can be easily written, as $F_Q$ simply projects onto $\bm{\alpha}$. In order to get $\tilde{F}_Q^{-1} \tilde{F}_Q = \bm{\alpha}^T \bm{\alpha} / \| \bm{\alpha} \|^2$ (identity on the image of $F_Q$), we must have
\begin{equation}
	\tilde{F}_Q^{-1} = \frac{ \alpha_i \alpha_j }{t^2 \| \bm{\alpha} \|^4}.
\end{equation}
$\bm{\alpha}^T \tilde{F}_Q^{-1} \bm{\alpha}$ is then equal to $1/t^2$, saturating the second inequality in Eq.~\eqref{eqn:multiparamqcrb} for the basis vector $\bm{b}$ corresponding to the largest $\bm{\alpha}$ component, $\alpha_b$= 1.

To evaluate the classical Fisher information in this case, we note that the final measurement \cite{Wineland1994} projects onto one of two outcomes with probability $\sin^2  \left( \adotq t/2 \right)$ and $\cos^2  \left( \adotq t /2 \right)$. Therefore the classical Fisher information is simply:
\begin{align}
	F &= \frac{\left(\pdpd{\left( \sin^2 \frac{\adotq t}{2} \right) }{\adotq} \right)^2}{\sin^2 \adotq t/2}  + \frac{\left(\pdpd{\left( \cos^2 \frac{\adotq t}{2} \right) }{\adotq} \right)^2}{\cos^2 \adotq t/2}  \\ &= t^2.
\end{align}
This Fisher information also implies the variance bound in Eq.~\eqref{eqn:ptebound}.

It may seem surprising that an optimal measurement can be one in which most qubits spend some of the measurement time idle. Since more time yields more signal, intuition suggests that the most effective strategy would make better measurements on the less-weighted qubits rather than keep them off for much of the measurement time. For example, by disentangling a qubit from the larger state halfway through the protocol, a separate measurement could be made on $\theta_1 + \half \theta_2$ and $\half \theta_2$, which appears to yield more information than just measuring the quantity of interest $\theta_1 + \half \theta_2$. 
This reasoning fails because there is no way to use information about $\theta_2$ to improve an estimate of $\theta_1 +  \half \theta_2$ without also knowing about $\theta_1$. Because we do not know about the individual parameters, only a measurement of the entire function is usable and our scheme is optimal in this case.
However, once we account for pre-existing knowledge about the parameter values (drawn from physically-motived estimates or less-precise previous measurements) our bound will instead apply in the regime of asymptotically many measurements $(M \gg 1)$ and in that setting our scheme will also saturate it \cite{Gill1995}. This is because the value of prior knowledge becomes increasingly low as we accumulate measurement data.

One advantage of this protocol is that an eavesdropper cannot learn the result of the network measurement by capturing a subset of the nodes' $\hat{\sigma}^x$ measurement results. This privacy can be shown by tracing out the first qubit in Eq.~\eqref{eqn:postptestate}, which leaves no phase information in the resulting mixed state.  
The central node can receive the measurement outcomes from all other nodes but keep its own secret, and no eavesdropper is able to extract information from the broadcasted results. This is true even if the central node's qubit is unweighted (i.e., $\alpha_i = 0$), which follows simply from the properties of the GHZ state. 
\subsubsection{Using Spin Squeezed States}
\label{sec:tdsss}
The perfect security of the GHZ state arises because obtaining the measurement result requires every qubit, but this also implies an extreme sensitivity to noise. This noise can be a serious problem for metrological applications \cite{Escher2011,Vidrighin2014}. Because the GHZ state decoheres faster than an individual qubit, the advantage provided by entanglement is nullified if the interrogation time of the qubit is limited by its coherence time \cite{Huelga1997}.  However, in many settings the time spent on a single measurement will be much shorter than the decoherence time, for instance, to gather data on short timescales. In these cases, GHZ states still provide a metrological advantage. Note that dynamical decoupling \cite{Viola1999} or quantum error correction \cite{Kessler2014a,Lu2015} could be used to lengthen the effective decoherence time in some cases.

In other situations, however, it may be that decoherence is the dominant concern. In these situations, the best strategy uses a highly-symmetric entangled state which is more robust to noise than the GHZ state \cite{Huelga1997}. Under dephasing, these states can still offer a constant factor improvement over unentangled metrology. In this section, we show that spin-squeezed states can also function as inputs to the partial time evolution protocol, and so may be good candidates for a sensor network operating in a situation where decoherence limits the interrogation time. Squeezed states are collective spin states which, due to entanglement, have reduced variance along one axis of the collective Bloch sphere at the cost of increased variance along an orthogonal axis \cite{Kitagawa1993,Wineland1992}. Recently, it has been shown that these states may allow Heisenberg-scaling measurements even without single-particle detection, which makes them very attractive for experimental implementations \cite{Davis2016a}. 

We consider a state whose overall spin vector is aligned along $+x$, such that $\expect{\hat{\sigma}^x_i} \approx 1$. We assume that the other spin components have zero expectation value, but that the variance of the collective spin projection $\hat{J}_y =  \half \sum_i \hat{\sigma}^y_i$ is decreased while the variance of $\hat{J}_z$ is increased. We quantify this effect through the spin-squeezing parameter $\xi$ \cite{Wineland1992},
\begin{equation}
	\xi = \sqrt{ \frac{ \Var \hat{J}_y}{N / 4}}.
\end{equation}

Suppose that we perform Ramsey interferometry on such a state \cite{Wineland1992,Wineland1994}. The protocol includes both partial time evolution $\hat{U}(t)$ and a final rotation pulse $\hat{R}_x \left( \frac{\pi}{2} \right) = \exp \left( - i \frac{\pi}{4} \sum_i \hat{\sigma}^x_i \right)$.
A final measurement is made of the total spin projection $\hat{J}_z$ after applying these operations: 
\begin{align}
	\expect{\hat{J}_z (t)} &= \expect{ \hat{U}^\dag (t) \hat{R}_x^\dag \left( \frac{\pi}{2} \right) \hat{J}_z (0) \hat{R}_x \left( \frac{\pi}{2} \right) \hat{U} (t) }, \\
	&= \half \expect{ \sum_{i = 1}^N \hat{\sigma}_i^x \sin \alpha_i \theta_i t + \hat{\sigma}^y_i \cos \alpha_i \theta_i t }.
	\label{eqn:jzpostevolve}
\end{align}
If we specify that this expectation is to be taken over a squeezed state with $\expect{\hat{\sigma}_i^x} \approx 1$ and $\expect{\hat{\sigma}_i^y} = 0$, then our signal will be sensitive only to $\adotq$ if each individual phase is small:
\begin{equation}
	\expect{\hat{J}_z (t)}_\mathrm{squeezed} \approx \half \sum_{i = 1}^N \sin \alpha_i \theta_i t \approx \frac{t}{2} \sum_{i=1}^N \alpha_i \theta_i.
\end{equation}
This shows that a squeezed state can be used for measurements of linear functions. The sensitivity can then be calculated just as in Eq.~\eqref{eqn:ptesensitivity},
\begin{equation}
	\Var Q = \left. \frac{ \Var \hat{J}_z(t)}{ \left( \partial \expect{\hat{J}_z (t)} / \partial q \right)^2} \right|_{q = 0} = \frac{ \Var \hat{J}_y }{t^2/4} =  \frac{N \xi^2}{t^2} .
	\label{eqn:ssssensitivity}
\end{equation}
We evaluate the sensitivity at $q = 0$ because we are interested in small signals. Partial time evolution with spin-squeezed input beats the standard quantum limit if $\xi \leq \| \bm{\alpha} \| / \sqrt{N}$. Note that there are $N$ components of $\bm{\alpha}$ and therefore $\| \bm{\alpha} \|$ will generally be of order $\sqrt{N}$ assuming that the moments of the field being measured are well distributed. Squeezed states can achieve squeezing proportional to $N^{-1/2}$ \cite{Wineland1992,Kitagawa1993}, which approaches the bound in Eq.~\eqref{eqn:ptebound} up to numerical prefactors not scaling with $N$.

Other highly-entangled states such as Dicke states also have metrological value in the presence of noise and could also serve as input states to partial time evolution with similarly favorable scaling \cite{Dicke1954,Holland1993a,Zhang2014,Campbell2009,Lucke2014}.

\subsection{Time-Independent Protocols}
In this section, we present two other possible measurement schemes for linear combinations of parameters. Both of these differ from the protocols of Sec.~\ref{sec:tdprotocols} because they prepare a particular state and then allow for free evolution during phase accumulation, rather than using pulses to evolve for an effective time of $\alpha_i t$ on qubit $i$. We will present time-independent schemes that begin with both a GHZ-like state and the spin-squeezed state. Note that these protocols rely on assumptions about the size of signals $\theta_i$ or the evolution time $t$.
\subsubsection{Using GHZ-like Input State}
We begin by defining a single-qubit state $\ket{\bm{\tau}}$, where $\bm{\tau}$ is a vector whose elements are $\tau_j = -1,0,1$:
\begin{equation}
	\ket{\bm{\tau}}  = \bigotimes_{j=1}^N \begin{cases} \ket{0} & \tau_j \neq -1 \\ \ket{1} & \tau_j =  - 1 \end{cases}.
\end{equation}
We then define the entangled state $\ket{\psi(\bm{\tau})}$ as
\begin{equation}
	\ket{\psi(\bm{\tau})} = \frac{1}{\sqrt{2}} \left( \ket{\bm{\tau}} + \ket{\bm{-\tau}} \right).
\end{equation}
This state can be understood as a general class that includes the GHZ state as the case $\tau_j = 1$ for all $j$. For every $\tau_j = -1$, spin $j$ is flipped relative to the GHZ state, while for every $\tau_j = 0$, spin $j$ is entirely disentangled.

In order to measure $\adotq$, we will evolve $\ket{\psi(\bm{\tau})}$ under the Hamiltonian in Eq.~\eqref{eqn:hamiltonian} and then measure the following observable $\hat{\Pi}(\bm{\tau})$:
\begin{equation}
	\hat{\Pi} ( \bm{\tau} ) = \bigotimes_{j} \left(\hat{\sigma}_x^j\right)^{\tau_j}.
\end{equation}
That is, we multiply the outcomes of the individual projective $\sigma^x$ measurements for each qubit which was originally entangled with the others ($\tau_j \neq 0$). It can be shown that probability distribution of this observable is
\begin{equation}
	P\left( \hat{\Pi} = \pm 1 \middle| \bm{\tau}, \bm{\theta} \right) = \begin{cases} \cos^2 \left( \bm{\theta} \cdot \bm{\tau} t/2 \right) & 1, \\ \sin^2 \left( \bm{\theta} \cdot \bm{\tau} t/2 \right) & -1 \end{cases}.
\end{equation}

To create a final protocol, we will now randomize the choice of $\bm{\tau}$, which in turn means we will randomly select both the initial state and the final measurement. An overall sensitivity to $\adotq$ can be realized if the probability distribution for every individual spin to be $\tau_j$ is given by:
\begin{equation}
	P(\tau_j) = \begin{cases} \frac{ \alpha_j ( \alpha_j \pm 1) }{2} & \tau_j = \pm1 \\ 1 - \alpha_j^2 & \tau_j = 0 \end{cases}. 
\end{equation}

By then summing over $P(\bm{\tau})$, we find that $\expect{\hat{\Pi}(\bm{\theta}, t)} = 1 - t^2 \left( \adotq \right)^2$ to lowest order in $t$. Since $\hat{\Pi}^2 = 1$, we can use the same approach as Eq.~\eqref{eqn:ptesensitivity} to find that the sensitivity for this measurement is $\Var Q = 1/t^2$, leading to the same sensitivity as the time-independent protocol.

\subsubsection{Using Spin-Squeezed States}
To implement a time-independent protocol that makes use of a spin-squeezed input state, we will actually use a two-part measurement protocol. First we will derive a general expression that applies to both parts, and then show how they can be combined. 

Much as in Sec.~\ref{sec:tdsss}, we will use the Heisenberg evolution of the total angular momentum along one axis to evaluate the final observable. We can begin with the result of Eq.~\eqref{eqn:jzpostevolve}, but with two alterations. First, rather than $\hat{U}$ representing a partial time evolution on each qubit, instead it will be the full time evolution operator $\hat{U} = \exp \left( -i t \sum \theta_i \hat{\sigma}^z_i  \right)$. Second, we will add an additional operator at the beginning of the protocol, which we write as $\hat{Q}(\bm{\eta})$:
\begin{equation}
	\hat{Q} \left( \bm{\eta} \right) = \bigotimes_{i=1}^N \hat{r}^i_z \left( \eta_i \right).
\end{equation}
Here, $\hat{r}^i_z$ is the single-qubit rotation about the $z$ axis. That is, we apply a qubit-dependent rotation about the $z$ axis before we begin the evolution. The final operator $\hat{J}_z (t)$ will be:
\begin{equation}
	\hat{J}_z(t) = \hat{Q}^\dag(\bm{\eta}) \hat{U}^\dag (t) \hat{R}_x^\dag \left( \frac{\pi}{2} \right) \hat{J}_z (0) \hat{R}_x \left( \frac{\pi}{2} \right) \hat{U} (t)  \hat{Q}(\bm{\eta}). 
\end{equation}
The effect of $\hat{Q}$ is to add an additional phase to the evolution, meaning the final value for $\expect{\hat{J}_z(t)}$ can be found by substituting the angles $\theta_i t + \eta_i$ for $\alpha_i \theta_i t$ in Eq.~\eqref{eqn:jzpostevolve}. As a result, we find that the final expectation value is:
\begin{equation}
	\expect{\hat{J}_z(t)} =  \half \expect{ \sum_{i=1}^N \hat{\sigma}^x_i \sin \left( \theta_i t + \eta_i \right) + \hat{\sigma}_i^y \cos \left( \theta_i t + \eta_i \right) }.
\end{equation}
By using the conditions that $\expect{\hat{\sigma}^x_i} \approx 1$ and $\expect{\hat{\sigma}^y_i} \approx 0$, we find that:
\begin{equation}
	\expect{\hat{J}_z(t)} \approx \frac{1}{2} \sin \left( \theta_i t + \eta_i \right).
\end{equation}

Now we introduce a two-step protocol. In the first step, we perform this sequence (prepare a spin-squeezed state, add qubit-dependent rotations, evolve, measure $\hat{J}_z$) with $\eta_i = \phi_i$, where $\cos \phi_i = \alpha_i$. We will call the quantity measured $\hat{J}^{+}_z$. Then, we repeat the process with $\eta_i = - \phi_i$, and call the resulting quantity $\hat{J}^{-}_z$. The expectation value of the sum of these quantities is:
\begin{align}
	\expect{\hat{J}^{+}_z + \hat{J}^{-}_z} &\approx \frac{1}{2} \sum_{i=1}^N \sin \left( \theta_i t + \phi_i \right) + \sin \left( \theta_i t - \phi_i \right) \\
	&= \sum_{i=1}^N \cos \phi_i \sin \theta_i t \approx \sum_{i=1}^N \alpha_i \theta_i t.
\end{align}
Here, as in Sec.~\ref{sec:tdsss}, we have assumed that the phases to be detected, $\theta_i t$, are small enough to make the small-angle approximation. 

In order to evaluate the sensitivity of this measurement, we look at the point of zero signal as in Eq.~\eqref{eqn:ssssensitivity}. At zero signal, $J^+_z + J^-_z$ gives $\sum \alpha_i \sigma_i^y$. It can be shown that $\Var \sum \alpha_i \sigma_i^y \leq 4 \Var J_y$, and so, by the same calculations used in Eq.~\eqref{eqn:ssssensitivity}, the variance is no more than $4 N \xi^2/ t^2$. Note, however, that this assumes that both $\hat{J}_+$ and $\hat{J}_-$ are measured for time $t$. For a fairer comparison, we can replace $t$ with $t/2$ so the time required for the two-step protocol is the same as for one time-dependent round. In this case, the sensitivity is no worse than $16 N \xi^2/ t^2$.

Interestingly, this two-step protocol requires only single-qubit operations once the initial squeezed state is created. This may make it a more tractable scheme for experimental realizations of quantum enhancements in measurements of linear combinations of parameters. 
\section{Entanglement-enhanced molecular NMR}
Many applications of entangled sensor networks may emerge as distributed entanglement becomes easier to achieve. In this section we focus on an application which may be viable in the near future: nanoscale nuclear magnetic resonance (NMR) as a form of molecular microscopy. NMR has long been used to investigate the chemical composition of molecular structures and perform medical imaging \cite{Mansfield2004}. The spatial resolution of NMR had been limited to a few micrometers until the recent advent of nitrogen-vacancy (NV) center magnetometers \cite{Mamin2013,Rondin2014,DeVience2015}. These magnetometers are sensitive to nanotesla magnetic fields with spatial resolution on the nanometer scale and can be used to image molecules or single proteins deposited on a diamond layer with embedded NV centers \cite{Lovchinsky2016,Lazariev2015,Perunicic2016}.  

Nanoscale NMR applications are a promising setting for entanglement-enhanced sensor networks. 
The electronic spin associated with an NV center in diamond can be operated as a two-level system whose free evolution results in the accumulation of phase dependent on the local magnetic field \cite{DeVience2015}. Because NV centers are useful platforms for quantum information processing, entangling protocols already exist and have been demonstrated experimentally \cite{Dolde2013,Dolde2014,Zu2014,Bernien2013}. Our protocol is particularly useful for studies of chemical or magnetic dynamics, such as Ref.~\cite{Wolfe2016}, because the measurement timescale may be much shorter than the decoherence time of the GHZ state, making our noise-free treatment applicable.

Linear combinations of spatially separated field values are interesting measurement quantities in nanoscale NMR. Reference~\cite{Arai2015} describes an imaging protocol which combines many different Fourier spatial modes, and Ref.~\cite{Lazariev2015} similarly combines many signals to perform molecular microscopy. These measurements could be performed more accurately using entangled NV sensors. In addition, our entanglement scheme can perform simple subtraction of the signal between two qubits. This allows common mode noise subtraction between a sensor qubit and another qubit exposed only to environmental noise. In general, even if a full GHZ state of all sensors is not feasible, smaller clusters of entangled sensors can still enhance sensitivity.

Entanglement-enhanced imaging of objects larger than single molecules may also be a fruitful area of research. 
An experiment detecting the firing of a single animal neuron with accuracy near the standard quantum limit has already been performed \cite{Jensen2016}, making exploration of techniques surpassing the limit a natural next step. Similar experiments could demonstrate an enhancement due to distributed entanglement in the near future. 

\section{Outlook}
We have presented measurement protocols for quantum networks which are useful for measuring linear combinations of parameters and developed a Heisenberg limit for the optimal estimation of linear combinations. Our protocol can be considered a generalization of entanglement-enhanced Ramsey spectroscopy, as in Ref.~\cite{Bollinger1996}, to the measurement of spatially varying quantities. In the future, we hope to search for further protocols and to remove the requirements of small signal or evolution time where we have imposed them. We identified magnetometry in general and nanoscale NMR in particular as candidate applications of our protocol, but we wish to stress our protocol's significantly broader scope. In particular, we expect that our protocol will be useful for measuring spatially varying quantities in contexts such as gravimetry \cite{DeAngelis2008,Debs2011}, spectroscopy \cite{Dinani2016}, and rotation sensing \cite{Ragole2016,Gustavson1997,Kolkowitz2016}. Note there is also no requirement that the parameters measured in a linear combination be of the same physical source. For instance, a sensor network could measure a linear combination of both electric and magnetic fields.

In general, our protocol can be applied in any setting where Ramsey spectroscopy can be applied if the quantity of interest is nonlocal. In addition, recent work \cite{Layden2017} indicates that spatial correlations in measurements may be a useful tool for noise-filtering and error correction in quantum sensors. 

Many schemes for quantum sensing rely on coherence in photonic, rather than atomic, degrees of freedom, such as spectroscopic microscopy \cite{Kee2004}. A recent manuscript, Ref.~\cite{Proctor2017} provides a general framework for treatment of sensor networks which is applicable to photonic systems and others.

We would like to thank P. Barberis Blostein, J. Borregaard, T. Brun, C. Caves, M. Cicerone, Z.-X. Gong, M. Hafezi, M. Oberthaler, J. Simon, V. Lekic, E. Polzik, K. Qian, and J. Ye for discussions. This work was supported by ARL CDQI, ARO MURI, NSF QIS, ARO, NSF PFC at JQI, and AFOSR. Z. E. is supported in part by the ARCS Foundation. 
\bibliography{library}{}
\end{document}